\documentclass[multicol,aps]{revtex4b3}

\draft
\twocolumn

\begin{document}

\title{The Complexity of our Curved Universe}

\author{V.G.Gurzadyan}

\address{ICRA, Dipartimento di Fisica, Universita di Roma La Sapienza,
Rome, Italy and Yerevan Physics Institute and Garni Space Astronomy
Institute,
Armenia\\
E-mail: gurzadya@icra.it}

\begin{abstract}
The properties of the Cosmic Microwave Background (CMB) radiation  must
be different in flat, positively and negatively
curved universes. This fact leads to a direct way of determining the
geometry of the universe.
The signature of the predicted effect of geodesic mixing, i.e., of
the `chaotic'
behavior of photon beams in negatively curved spaces peculiar to Anosov
systems with strong statistical properties, has been detected
while studying the COBE-DMR 4-year data \cite{GT}. This possible
observation of the negative curvature of the universe suggests the need
to search for more effective ways to analyze the CMB data
expected from forthcoming high precision experiments.
Information theory offers such a descriptor
for the CMB
sky maps --- the Kolmogorov complexity --- as well as provides
novel insight into the problem of the loss of information and time
asymmetry in a hyperbolic universe.
\end{abstract}
\maketitle

\section{Introduction}

Cosmologists believe that CMB photons have been moving freely during
most of the lifetime of the universe.
If so, then reaching us without any modification since the
epoch of last scattering, as believed, the CMB photons
must contain information about the early phases of the universe.
However, given the enormous distances in space-time over which
the photons have traveled,
they must also carry certain imprints of the
expansion, geometry and even of the topology of the universe.

The expansion of the universe is reflected in the reddening of the CMB
photons, which concerns individual photons. On the other hand a bundle
of photons in a beam
can also contain information about the geometry of 3-space, i.e.,
can possess different properties depending on the curvature of the
universe.
Obviously, these properties have to be statistical ones if they concern
beams and not individual photons. The present technique of
CMB measurements
also deals with statistical characteristics, like the temperature
anisotropy amplitude $\Delta T (\theta)/T$, the angular autocorrelation
function, the anisotropy distribution over the sky, etc., and
one can expect that this data must contain information about the
geometry of space.

Here we review several CMB effects which will occur
in a negatively curved $k=-1$ Friedmann-Robertson-Walker (FRW) universe.
This curvature choice is motivated not only by the fact that the
low density and negatively curved FRW universe is consistent with
or favored by various observational data;
see e.g. \cite{Huc}$^{,}$ \cite{Coles}. It appears likely that
it can also provide alternative insight into such basic problems as the
second law and the arrow of time.

The free motion of photons is described by null geodesics. The motion
of photon beams is therefore given by a geodesic flow. The remarkable
properties
of geodesics in hyperbolic spaces have been continually attracting
attention
over many decades \cite{Had}$^{-}$\cite{Hopf}. In particularly it was
shown that a geodesic flow on compact manifolds with constant negative
curvature possesses strong statistical properties, i.e., according to
the
classification of the theory of dynamical systems it possesses the
spectral
properties of ergodicity, mixing of all degrees and K-mixing. Moreover,
it was
proven to be isomorfic to the Bernoulli shift, and to have positive
Kolmogorov-Sinai (KS) entropy, exponential decay of the time correlation
functions, and countable Lebesgue spectrum of conjugated
group of unitary operators, and is also a structurally stable (coarse)
system,
i.e., is an Anosov system \cite{Ano}.

Considering the behavior of geodesics in a hyperbolic FRW universe,
i.e., the
effect of geodesic mixing, one arrives
at the following consequences for the observable statistical parameters
of the CMB \cite{g1}$^{-}$\cite{book}:

(1) decrease of the amplitude of temperature anisotropy in time;

(2) flattening of the angular autocorrelation function, i.e.,
independence of the sky angle;

(3) distortion of anisotropy sky maps.

The rate of the damping, flattening and distortion is determined by
the KS-entropy, which itself depends on the curvature or the diameter of
the universe ---
the only scale parameter in a homogeneous isotropic space.
Evidently nothing is happening to any individual photon during free
propagation after the last scattering epoch, and these effects are
purely
statistical ones and are determined by the principal limitations of
obtaining information during measurements,
i.e., by the  impossibility of
reconstructing the trajectory of any given
photon while observing within a finite smoothing angle due to the
overlapping
of exponentially deviating geodesics in any cut of phase space.

Consider the third effect. Actually, it provides a way to trace the
geometry
using the CMB sky maps, since the effect is absent for $k=0, +1$
curvatures.
The predicted distortion \cite{g2} can be roughly attributed to an
elongation of anisotropy spots depending on the value of the curvature
of the universe, and hence on the density parameter $\Omega$, as well as
on the redshift of the last scattering epoch, i.e., the distance covered
by the photons while moving along geodesics.
Such a study has been performed using the COBE-DMR 4 year data and
special pattern recognition codes, and
a statistically significant signal of the sought after effect has been
detected \cite{GT}. If it is due to the effect
of geodesic mixing, as predicted, then one has
a direct indication of the negative curvature. Forthcoming more accurate
observations will be helpful for confirmation of that result. Moreover
this effect may also provide a way to determine the rate of
expansion;
this has become an interesting topic due to the recent claims on its
possible accelerating expansion rate \cite{Peeb}.

We have to note, however, that talking about the geometry of the 3-space
as
determined by geodesic mixing and CMB data by no means implies that
one can also have unambiguous information about the topology of the
universe.
Indeed, the Einstein equations define the geometry but not the
topology of the space. The same geometry
can correspond to different topologies, e.g. the flat $k=-0$ space can
have $R^3, R^1 \times T^2, R^3 \times S^1, T^3$ and other
topologies. The same is true for $k=-1, +1$ geometries.
This includes both closed spaces,
i.e., compact and without boundary, like $S^3, T^3$,
k-handled tori, etc., as well as non-compact spaces.
Therefore the study of the topology of the universe
using CMB data is another intriguing problem, enabling one
to obtain constraints and make efficient predictions \cite{Top}.
The study of CMB properties in open models started in\cite{Wil}
is  complicated due to the difficult problem of eigenfunctions of differential
operators (Sobolev problem) in hyperbolic spaces when their topology
cannot be defined a priori.   

The geodesic mixing and CMB in a hyperbolic universe thus reflects
the inevitable loss of initial information peculiar to chaotic dynamical
systems. Therefore, this problem can be studied also from the point of
view of information theory.
Within that approach we show the efficiency of a new tool,
the Kolmogorov (algorithmic) complexity \cite{Kolm}--\cite{Cha}, for the
study of the CMB data, since this complexity is related
to the curvature of the universe \cite{G}.

\section{Geodesics in (3+1) and (3)-spaces}

First let us consider the null geodesics, that is the free motion of
photons.
We will show how the null geodesics in  a (3+1)-manifold can be
projected into its 3-hypersurface, so that the projection will be a
geodesic of the latter. More details can be found in \cite{book}.

Consider a 3-dimensional Riemannian manifold $U$ and a (3+1)-dimensional
Lorentzian manifold $W$. Let the metric of the latter be $^4{\bf g}$ and
let $W$ be oriented and time--oriented. Let
\begin{equation}
  \imath_t:U\to W,
\end{equation}
be an embedding of $U$, such that the embedded manifold
\begin{equation}
\Sigma_t=\imath_t(U)
\end{equation}
is a space-like hypersurface in $W$,
and the induced metric $\imath^*(^4{\bf g})={\bf ^3g}$ is also a
Riemannian
metric on $U$.

A one-parameter family of vector
fields ${\bf Y}_{U_t}$ on the embedded hypersurfaces $U_t\equiv
\imath_t(M)$
is defined via the derivative
\begin{equation}
  {\bf Y(t)}\equiv \frac{d\imath_t}{dt} .
\end{equation}
It can be split into normal and tangential components to $\Sigma_t$
\begin{equation}
{\bf Y}=N{\bf Z}_U + {\bf X} ,
\end{equation}
where ${\bf Z}_U$ is a time-like normal to $\Sigma_t$.

The Riemannian metric $^4{\bf g}$ of $W$ can be represented in the form
\begin{equation}
{\bf g}
  =-N^2{\bf dt}\otimes{\bf dt}
  +\,^3g_{ab}({\bf dx}^a+X^a{\bf dt})\otimes({\bf dx}^b+X^b{\bf dt}) ,
\end{equation}
where $\,^3g_{ab}=(g_t)_{ab}$ is the metric of $U$ and
\begin{equation}
{\bf g}_t=\imath_t^{*4}{\bf g} .
\end{equation}

By means of the vector field ${\bf Y}$ one can define uniquely  the
projection
\begin{equation}
\pi : W \rightarrow U ,
\end{equation}
which obviously depends on $N$ and ${\bf X}$.
As a result any curve
$\gamma$ in $W$ can be projected on a curve $c$ in $U$.
$$
\begin{array}{cc}
   \begin{array}{rcl}
                W & \stackrel{\textstyle \pi}{\rightarrow}   & U\\
  \gamma \uparrow &                                          & \uparrow
c\\
                R & \stackrel{\rightarrow}{\textstyle\lambda} & R
   \end{array}
 &\qquad
          \begin{array}{ll}
             &c=\pi\circ\gamma\circ\lambda .
          \end{array}
\end{array}
$$
Now can one find out what conditions
$^4{\bf g},N,{\bf X},\phi$ should satisfy in order that the  projection
of
any null geodesic on $W$ be a geodesic on $U$ with respect to its metric

$^3{\bf g}$?

The necessary conditions can be shown to be \cite{LMP}
\begin{eqnarray}
&&N=1,\\
&&{\bf X}=0,\\
&&{\bf g}=a^2(t)\cdot{\bf h} ,\\
&&\lambda(t)=A \int^t a^{-1}(s)ds ,
\end{eqnarray}
where ${\bf h}$ is one of the metrics of the maximally symmetric
homogeneous-isotropic $3$-manifold.
We are interested in such metrics since are considering the FRW
universe;
for non-maximally symmetric metrics the problem of obtaining
analytical relations for necessary and sufficient conditions
is more complicated.

For time-like trajectories of non-zero mass particles,
compare the corresponding relations for isotropic and time-like
trajectories:
\begin{equation}
\lambda(t)=\left\{
\begin{array}{lr}
 A \int^ta^{-1}(s)ds
                   &\hbox{isotropic}\\
 B \displaystyle{\int^t a^{-1}(s)[C^2+a^2(s)]^{-1/2}ds}
                   &\hbox{time-like} ,
\end{array}
\right.
\end{equation}
where $A, B, C$ are constants.
Their difference is responsible for the different efficiencies of the
geodesic mixing for photons and non-zero mass particles. Namely,
in the former case the characteristic  time scale of the effect is
much smaller, so that photons have time to mix, while the matter does
not.

\section{The Effect of Geodesic Mixing}

A geodesic flow on the ($3+1$)-manifold $W=U\times R$ with
Robertson--Walker metric thus can be reduced
to a geodesic flow on the ($3$)-manifold $U$
with metric $a_0^2{\bf h}$ and affine parameter $\lambda$.

The behavior of close geodesics is determined by the equation of
geodesic
deviation --- the Jacobi equation
\begin{equation}
  \frac{d^2{\bf n}}{d\lambda^2}+k{\bf n}=0,\label{J}
\end{equation}
where ${\bf n}$ is the deviation vector and $k=0,-1,+1$ is the
normalized
curvature.
This equation has the following solutions depending on the three values
of the curvature $k$
\begin{equation}
  {\bf n}(\lambda)=\left\{
   \begin{array}{ll}
     {\bf n}(0)+\dot{{\bf n}}(0)\lambda                  & k=0 ,\\
     {\bf n}(0)\cos\lambda+\dot{{\bf n}}(0)\sin\lambda   & k=+1 ,\\
     {\bf n}(0)\cosh\lambda+\dot{{\bf n}}(0)\sinh\lambda & k=-1 .
   \end{array}
   \right.\label{sol}
\end{equation}
Thus in the case of negative constant curvature, nearby geodesics
in $U$ deviate by an exponential law
\begin{equation}
    l(\lambda)=l(0)\exp(h\lambda) ,
\end{equation}
where $h$ is the KS-entropy and in accordance with the Pesin theorem
equals the sum of the positive Lyapunov numbers
\begin{equation}
h=2a^{-1} ,
\end{equation}
where $a=\sqrt -R$ is the diameter of the universe, $R$ is the
curvature,
and $h=0$ when $k=0$ or $k=+1$ (cf. \cite{ET});
this is also valid for non-compact spaces.
Taking into account the expansion $a(t)$ (from its initial value
$a(t_{0})$), the true deviation in $W$ will be
\begin{equation}
   L(t)=L(t_0)\frac{a(t)}{a(t_0)}\exp(h\lambda(t)) .
\end{equation}
The next crucial point
concerns the behavior of the time correlation functions for a geodesic
flow.
As it was proved by Pollicott \cite{Pol}, for 3-spaces the time
correlation
function of a geodesic flow $f^t$ decreases by an exponential law,
i.e., $\exists c>0$ such that for all
$A_1, A_2 \in L^2(SM)$
\begin{equation}
\bar b(t) = \mid \int A_1 (f^tu) A_2(u)d\mu - \int A_1(u)d\mu(u)
\int A_2(u)d\mu(u)\mid=
\end{equation}

\begin{equation}
const \parallel A_1\parallel \parallel A_2 \parallel
\dot (1+t) e^{-h(f)t} + O(e^{-ht}) ,
\end{equation}
where $\mu(SM)=1$ is the Liouville measure and
\begin{equation}
\parallel A \parallel = [\int A(u)^2 d\mu(u)]^{1/2} ,
\end{equation}
and $h$ is again the KS-entropy.

Two properties of Anosov systems have particular importance for our
problem.
First, the Anosov systems are {\it coarse} or {\it structurally stable}
systems, i.e., they are topologically equivalent to any sufficiently
close dynamical
system \cite{Ano}.
In other words a perturbed Anosov system also has the
properties of an Anosov system.  Although we live in
a perturbed and not exact FRW universe, in view of the structural
stability of Anosov systems, the geodesic flows in
homogeneous-isotropic 3-hypersurface with small perturbations of the
curvature
have to possess the properties of an Anosov system as well.

The second property is the homogeneous mixing of Anosov systems. This
means
that in time the exponential stretching and contracting in equal numbers
of dimensions (due to the Liouville theorem) of the initial
configuration
in the course of evolution will tend to become distributed homogeneously
over all coordinates of the phase space. This property is responsible
for the appearance of a complex shape of the anisotropies
\cite{g2} (Figure 1).










Moreover, this property has to be independent of the temperature
threshold, thus enabling it to be distinguished from similar effects
caused by noise \cite{Bond}. Just
the threshold independence of the elongated structures
has been detected in \cite{GT}, indicating the reality of the sought
after effect of the curvature.\footnote{Note, that the
map distortion due to geodesic mixing has no direct relation with the
effects discussed in \cite{Bar}, and exists whatever the initial
shapes of spots at the last scattering surface are.}

For the last scattering redshift $z$
and the present value of the density parameter $\Omega$ the
exponential
factor can be written in the form \cite{book}
\begin{equation}
e^{ht}= (1+z)^2[1+\sqrt{(1-\Omega})/(\sqrt{1+z\Omega} +
\sqrt{1-\Omega_0})]^4 .
\end{equation}
In particular 
{\it the measured CMB temperature will tend exponentially to a constant
mean temperature by time: the isotropy is the limiting state}
\begin{equation}
\lim_{t\to \infty} T_{\lambda} =\bar{T} .
\end{equation}
For the normalized temperature autocorrelation function
\begin{equation}
 {\cal C}_{\lambda}(\theta,\beta)
     =\langle T_{\lambda}(u) T_{\lambda}(v) \rangle_{g(u,v)
     =cos\theta}\ ,
\label{C}
\end{equation}
the following inequality holds
\begin{equation}
  \left|{\cal C}_{\lambda}(\theta,\beta)-1\right|
  \leq  const \cdot \left|{\cal C}_0(\theta,\beta)-1\right|
          \cdot\frac{1}{(1+z)^2}
          \cdot\left[\frac{\sqrt{1+z\Omega}+
          \sqrt{1-\Omega}}{1+\sqrt{1-\Omega}}\right]^4 \ ,
\end{equation}
where $\theta$ and $\beta$ are the sky (separation) and observing
beam angles, respectively.
Thus   {\it the autocorrelation function $C(\theta,\beta)$
tends to become constant with respect to the sky angle $\theta$ in
time, regardless of its form at the last scattering
surface.}

  Numerically, the exponential factor, for say
$\Omega\simeq 0.2$ and expansion rate $t^{\alpha}$, $\alpha=2/3$, is
\begin{equation}
   \exp(ht) \simeq 10^{-3} .
\end{equation}
Using the fact that the geodesic flow is an Anosov system,
it can also be shown \cite{g3} that there exists an angle $\phi$ such
that
the smoothing factor $s$ of $\delta T/T(\beta)$ is
almost constant if either $\beta\gg\phi$ ($s\sim e^{-h\lambda}$)
or $\beta\ll\phi$ ($s\sim 1$), and it increases as
$\beta$ decreases at  $\beta\sim\phi$ by the law
$s\sim const/\beta\cdot e^{-h\lambda}$.
The appearance of the {\it
smoothing angle} $\phi $ is due to the fact that CMB measurements
include
averaging within some beam angle and time period,
i.e., statistical smoothing with inevitable loss of information.
Therefore, the more narrow
is the beam size, the less information is lost in the smoothing
in terms of the temperature autocorrelation function. Maximal
information, e.g. the anisotropy corresponding to exactly the last
scattering surface, should be obtained while measuring by beams within
some limiting angle. Measurements on smaller
angles should not influence the autocorrelation function which should be
determined by physical conditions at the last scattering surface
(its thickness, the Silk effect, etc). The predicted increase of the
anisotropy at small beam angles for the parameters mentioned above is
close to the angular region of the Doppler peak,
therefore the separation of these effects can be of particular
importance.

The value of the KS-entropy is therefore determining the observational
effects of the loss of information.
If for CMB photons the exponential factor takes values up to $10^3$, for
photons
arriving from redshifts of quasars or galaxies, it is sufficiently
smaller: $(1+z) \cdot 10^{-3}\sim5 \cdot 10^{-3}$, i.e., the latter
photons
have little time to feel the geometry.

The effect of geodesic mixing with positive KS-entropy
will also mean that certain properties of the CMB can seem
similar on scales larger than the scale of the horizon not because of
exchange of information but due to a  corresponding loss of information
about the initial conditions.
In the next section we will deal with the information theory approach
to the CMB problem.

\section{Kolmogorov Complexity}

Various descriptors have been
proposed to extract the cosmological information from the CMB data.
Particularly for CMB maps those descriptors include, for example, the
hot spot number density, genus, correlation function of local
maxima, Euler-Poincare characteristic, percolation, wavelets etc.
(see e.g. \cite{Silk}$^{,}$ \cite{Fang}).
Most of these descriptors have been already applied to the analysis of
COBE-DMR sky maps \cite{smoot}$^{,}$ \cite{Bennett}.

As discussed above, the geodesic mixing will lead to complex
structures on the CMB sky maps \cite{book}.
To describe quantitatively the latter we suggest using the
invariant definition of complexity ---
Kolmogorov or algorithmic complexity --- introduced in the 1960s by
Kolmogorov,
\cite{Kolm}, and independently by Solomonoff and Chaitin (see \cite{Ch},
\cite{Cha}).
The efficiency of the concept of complexity for the basics of
classical and quantum physics and the second law of thermodynamics have
been studied by Zurek, Caves, Bennett and others \cite{Zu}$^{,}$ \cite{Time}.

The Kolmogorov complexity $K_u$ is defined as the minimal length of the
binary coded program (in bits) which is required to describe the system
completely, i.e., that will enable one to recover the initial system via
a given computer. The complexity of an object $y$ at a
given object $x$ is defined as
\begin{equation}
K_{\phi(p)}(x) =min_{p:\phi=x} l(p) ,
\end{equation}
where $l(p)$ is the length of the program $p$ with respect to the
computer $\phi(p)$
describing the object completely, i.e., at the $0-1$ representation:
$l(\O)=0$. The fundamental feature of Kolmogorov's formulation is the
independence of complexity on the computer which has to halt upon
running the
program.
A computer is considered `universal' if for any computer $\phi_C$
there exists a constant $S_C$ which can be added to any program $p$,
so that $\phi_Cp$ should execute the same operation on computer $\psi$
as the program $p$ on computer $\phi$.
Therefore the Turing machine
can be considered as a universal computer while computing the
complexity.

The complexity is closely related with another
basic concept --- random sequences. The most general definition by
Martin-L\"{o}f \cite{Martin}  formalizes the idea of Kolmogorov
that random
sequences have a very small number of rules compared to their lengths;
the rule
is defined as an
algorithmically testable and rare property of a sequence. Indeed, the
properties of complexity and randomness
are not completely the same, although they are
closely related for typical sequences \cite{ZL}.
Therefore in our problem, in principle, the estimation of the randomness
of the data string (digitized figure) has
to correlate with the estimation of the complexity; we will not discuss
the randomness here but it could be an interesting problem for future
studies.

It has been proven that
no shortest algorithm exists enabling one to decide whether a given
complex-looking sequence is really complex \cite{Cha}$^{,}$\cite{ZL}.
The complexity is the amount of information which is
required to determine uniquely the object $x$. Kolmogorov had proved
that
the amount of information of $x$ with respect to $y$ is given by the
formula \cite{Kolm}
\begin{equation}
I(y:x)=K(x)- K(x\mid y) ,
\end{equation}
where the conditional complexity
\begin{equation}
K(x\mid y)=\hbox{min}\, l(p) ,
\end{equation}
is the minimal length of the program required to describe the object $x$
when
the shortest program for $y$ is known.
Note, that
$K(x\mid x)=0$ and $I(x:x)=K(x)$ and $I(x:x)\succeq 0$;
where $A \succeq B$ denotes $A\leq B + const$.

The complexity measured in bits is related to KS-entropy via
the relation \cite{Cav}
\begin{equation}
\label{I}
\Delta I =\log_2 (2^{h(f^t)(t-t_0})=h(f^t)(t-t_0) ,
\end{equation}
where the loss of information $\Delta I$ during the time interval
$t-t_0$ is
\begin{equation}
\label{K}
\Delta I = K_u(t) - K_u(t_0) ,
\end{equation}
i.e., the information corresponding to the distortion of the pattern
from
the initial state $t_0$ (the last scattering epoch) up to the observer
at $t$ i.e., at $z=0$.
This is guaranteed by the Shennon-McMillan-Breimann theorem \cite{Shen}
stating the
uniform exponential rate of loss of information in Anosov systems.

If the length of a sequence $x$ is $N$ then the obvious upper
limit can be established
\footnote{If $x$ is the binary representation
of some integer $N_0$, then $N\approx \log _2N_0$.}
\begin{equation}
\label{1}
K_U(x)<N .
\end{equation}
Let us estimate the fraction of the sequences among all
N-bit sequences for which
\begin{equation}
K_U(x)<N-m .
\end{equation}
This means that there exists a program of length $N-m$ which computes
$x$.
The total number of such programs of such a length cannot be larger than
$2^{N-m+1}$; this is the upper limit without taking into account the
prefix-free condition, i.e., the programs cannot be a prefix for some
other
program.
\footnote{A word $a$ is called prefix for a word $b$ if $b=ac$ with some
other word $c$.}
 Thus, we have the following upper limit
\begin{equation}
(2^{N-m+1}-1)/ 2^{-N}\approx 2^{-m+1} .
\end{equation}
This value is small if $m$ is sufficiently large. Thus a more general
relation than (\ref{1}) can be established
\begin{equation}
K_U(x)\approx c(x)\, N, \ c(x)\approx 1 .
\end{equation}

Return again to the cosmological problem. In a $k=-1$
FRW universe any initial CMB structure observed
at redshift $z_{obs}$ should have a more complex shape
than at $z < z_{obs}$
\begin{equation}
K(CMB\, spot\mid z=z_{obs}) > K(circle\mid z< z_{obs}) ;
\end{equation}
here `circle' denotes a circle with Gaussian or other fluctuations.
The complexity estimated for the anisotropy spot observed now
should exceed the complexity of the primordial spot (in addition to
the scale expansion by the factor $1+z$)
by an amount depending on the KS-entropy and, hence on the curvature
and the distance of the last scattering surface.

From Eqs.(\ref{I}) and (\ref{K}) we come to a simple equation
linking the geometry with the complexity
\begin{equation}
\Delta K = (2/a) \Delta {\cal T} .
\end{equation}
i.e., the relative complexity of the observed spot with respect
to a circular spot (with Gaussian or so fluctuations which one can
expect
for flat or positively curved spaces) will
determine the curvature of the hyperbolic universe, where $\Delta \cal
T$ is the
time elapsed since photons started to move freely and thus
tracing how curved the 3-D space is.

Although in general the shortest program cannot be reached,
i.e., the exact complexity cannot be calculated, in certain problems
the results obtained cannot be too far from that value.
The calculation of relative complexity of a perturbed and non-perturbed
object by means of a given computer and developed code (although the
latter
cannot be proved to be the shortest possible), has to reflect the
complexity introduced by the perturbation. In our problem the
complexity is the `measure the perturbation' caused by the curvature.

The cosmological information should be extracted from the CMB maps in
the
following way. First, certain criterion should define the spots
represented
via given configurations of pixels (e.g.\cite{GT}).
Then, the computation of the
complexity should be performed by means of the length of a special
compressed
code (string) completely defining the spots. Since the basic program
will
be the same,
the only changes will be due to different data files, i.e., the
coordinates
of the pixels of various spots. This problem was technically solved in
\cite{AGS} and is discussed in the next section.

\section{Complexity: Algorithm for CMB maps}

We shall now describe an algorithm for estimating
the complexity of spots given by certain pixel configuration on a grid
and present the results of computations for a series of structures
of different complexity, i.e., demonstrate the calculability
of such an abstract descriptor as the Kolmogorov complexity for the CMB
digitized maps. The correlation of complexity $K$ of the anisotropy
spots
with their Hausdorff (fractal) dimension $d$ will be shown as well.
We observe the correlated growth of the complexity $K$ and $d$ with
the increase of the complexity of the geometrical shape of the spots,
starting from the simplest case --- the circle.

To develop the algorithm of estimation of complexity one should clearly
describe in which manner the objects, namely the anisotropy spots,
are defined.
The COBE-DMR CMB sky maps have the following structure \cite{smoot},
\cite{Bennett}.
They represent a $M\, \times N$ grid with pixels determined by the beam
angle
of the observational device; more precisely the pixel's size
defines the scale
within which the temperature is smoothed, so that each pixel is
assigned a certain value of the temperature (a number).
For example, COBE's grid had 6144
pixels of about 2.9$^{\circ}$ size each, although they do not
uniformly contain information about the CMB photons. By `anisotropy
spots'
we understand the sets of pixels at a given temperature threshold
\cite{To}.

Our problem is to estimate the complexity of the anisotropy spots, i.e.,
of various configurations of pixels on the given grid: the size of
the grid, and both the size and the number of pixels are crucial for the
result.

We proceed as follows. Each row of the grid is considered as an
integer of $M$ digits in its binary representation, `0' corresponding
to the pixels not belonging to the spot, and `1' to those of the spot.
Considering all $N$ rows of the grid in one sequence (the second row
added
to the first one from the right, etc.) we have a string of length $N
\times
M$ in binary form with complexity $K$.

Strictly speaking we can estimate only the upper limit of $K$
corresponding
to a given algorithm. By  algorithm we understand
the computer
program (in PASCAL in our case), along with the data file,  describing
the
coordinates of the pixel of the spot.  Namely the data file includes
compressed information about the string of digits.
The program is a sequence of commands performing reconstruction of the
string
and calculations of the corresponding lengths.
The complexity of the figure will
be attributed to the file containing the information aobut the position
of the pixels.

The code describing the spot works as follows. As an initial pixel we
fix the upper left pixel of the spot and move clockwise along its
boundary.
Each step  --- a `local step' ---  is a movement from a current pixel to
the
next one in the direction given above. This procedure is rigorously
defining the
`previous' and `next' pixels. Two cases are possible. First, when
the next pixel (or several pixels) after the initial one is in the same
row: we write down the number of pixels in such a `horizontal step'.
The second case is, when the next pixel is in the vertical direction;
then we perform the local steps in the vertical direction (`vertical
step') and
record the number of corresponding pixels. Via a sequence of
horizontal and vertical steps we obviously return to the initial pixel,
thus defining the entire figure via the resulting data file.

Obviously, the length of the horizontal step cannot exceed the number of
columns, i.e., $N$, while the vertical step cannot exceed $M$, requiring
$\log_2M$ and $\log_2N$ bits of information, correspondingly.
For the configurations we are interested in, the lengths of the
horizontal and vertical steps, however,
are much less than $\log_2M$ and $\log_2N$ and therefore we need a
convenient code for defining the length of these steps. Our code is
realized for $M=N=256$; apparently for each value of $M$ and $N$ one has
to choose the most efficient code.

After each step, either horizontal or vertical, a certain number of
bits of information is stored. The first two bits will contain
information
on the following bits defining the length of the given step in a manner
given in the Table 1.
\begin{table}
\centering
\caption{}
\medskip
\begin{tabular}{lll}
\hline
\hline
first 2 bits & next bits   & \\
\hline
0 1          & 1           & \\
1 0          & 2           & \\
1 1          & 3           & \\
\hline
\end{tabular}
\end{table}
The case when the first two bits are zero, denotes: if the following
digit
is zero than the length of the step is $l_s=0$, and hence no digits of
the same step do exist; if the next digit is $1$, then 8 bits are
following,
thus
defining the length of the step. If $l_s=1$, then after the combination
$0\, 1$ the following digit will be either $0$ or $1$ depending
whether the step
is continued to the left or to the right with respect to the direction
of the previous step. When $l_s=2$ or $3$, after the combination $0\, 1$
the file records $0$ in the first case, i.e., $l_s=2$, and $1$ in the
second. When $l_s=4,...,7$, then after the combination $1\, 1$ the file
records $0$ and $1$, at the left and right steps, and after two digits
in binary form of the step length $l_s=3$. Finally, when $l_s>8$, the
combination $0\, 0\, 1$ is recorded, followed by the 8 bits of the step
length $l_s$ in binary representation.

Thus, all possible values of the step length $l_s$ (they are limited
by $M=N=256$) are taken into account and the number of bits attributed
to
the length in the file depends on $l_s$ in the manner shown in Table 2.
The figure recorded in the data file via the described code can be
recovered unambiguously without difficulties.

\begin{table}
\centering
\caption{}
\medskip
\begin{tabular}{lll}
\hline
\hline
step length  & bits & \\
\hline
0, 1          & 3    & \\
2, 3          & 4    & \\
4-7           & 5    & \\
 8           &11    & \\
\hline
\end{tabular}
\end{table}

Obviously one cannot exclude the existence of a code compressing
more densely the information about the pixelized spots, however, even
these codes appear to be rather efficient. Namely, the length of the
program recovering the initial figure from the stored data file is 4908
bits, and it remains almost constant at the increase of $N$ and $M$.


\section{Hausdorff dimension of CMB spots}

The association of local exponential instability and deterministic
chaos with fractals is well known (see e.g. \cite{ZS}). Hence the
idea to estimate the Hausdorff dimension of the spots is natural.
We recall that the Hausdorff dimension is defined as the limit
\begin{equation}
d_H=\lim_{\varepsilon\to 0}\frac{\ln N(\varepsilon)}{\ln
(1/\varepsilon)} ,
\end{equation}
where $N(\varepsilon)$ are circles of radius $\varepsilon$ covering at
least one
point of the set. By the definition of Mandelbrot the set is fractal if
the
Hausdorff dimension exceeds the topological  dimension.

To compute the Hausdorff dimension we used the code {\it Fractal}
by V. Nams \cite{Program}, but its application was not so
straightforward.
The main problem to be solved was the approximation of the boundary of
the
pixelized figure via a smooth curve,
so that its Hausdorff dimension can be determined by the above mentioned
code.
The trivial consideration of the profile of the pixels, obviously would
introduce artificial fractal properties to the
spot as a result of instrumental nature of pixel sizes.
We used the following
procedure: the centers of three or more neighbor pixels were connected
by a
line and its distance $h$ from the centers
of the intermediate pixels has been calculated (it is obviously zero if
the pixels are in one row).
 If $h$ exceeds some chosen value, namely 0.5 of the size
of the pixel, then the line was adopted as a good approximation of the
boundary
curve of the pixels. Otherwise,
the centers of the next pixels are involved, etc.
The runs of test (trivial) figures with various values of $h$ showed the
validity of this procedure.
















Figure 2 shows the results of computations of complexity and Hausdorff
dimension for a sequence of toy spots, starting from a circle.

\section{Discussion}

Thus the hyperbolicity of the universe has to be reflected in the
properties
of CMB. The motion of photons in a $k=-1$ FRW universe via the
geodesic flows/Anosov systems is linked with deterministic
chaos and loss of information occurring for negative curvature.

Then the information theory approach can be applied.
Kolmogorov (algorithmic) complexity can be an efficient descriptor of
CMB sky maps: we presented an algorithm for its computation
for a given configuration of pixels on a grid.
We showed that it is calculable for CMB maps, if
we are interested in the
relative complexity $K_i-K_1$ (or $(K_i-K_1)/K_1$); for details see
\cite{AGS}.

Together with the previous results on the rate of exponential mixing
of geodesics determined by the Kolmogorov-Sinai (KS) entropy, which
itself is related by the diameter (curvature) of the universe,
this provides a new informative way of analyzing the
CMB data.

Although the COBE-DMR data reveal certain genuine spots
\cite{Tor}$^{,}$\cite{Cay}, they were not sensitive enough for reliable
evaluation of the shape of the spots.
The next generation of observations, such as the Planck Surveyor and
MAP, should allow one to apply the technique described here.

The chaos/information approach to the CMB problem briefly discussed
above
can open ways
for even more general insights. Namely, the effect of geodesic mixing
can be only one of the manifestations of a much deeper link with the
fundamental
physical laws --- {\it negative curvature -- mixing -- second law of
thermodynamics -- arrow of time}.
Thus, {\it we observe the time asymmetry and the second law because
we live in a universe with negative curvature, and those laws may not
be the same in a flat
or positively curved space}. These aspects, including the
relations between thermodynamic and cosmological arrows of time,
are discussed in a separate paper \cite{AG}.

The complexity and information way of thinking can be valuable also
in other cosmological problems.

I am grateful to R. Penrose and W. Zurek for valuable discussions, and
my collaborators A. Kocharyan, S. Torres, A. Allahverdyan and
A. Soghoyan for our joint work on these interesting problems and
R.Jantzen for many useful comments.

\end{document}